\documentclass{aa}  

\usepackage{longtable}
\usepackage{graphicx}
\usepackage{txfonts}
\usepackage{longtable,lscape}

\def\sun{\hbox{$\odot$}}
\def\R23{\mbox{$\rm R_{23}$}}

\def\arcsec{\hbox{$^{\prime\prime}$}}

\def\kmsmpc{km s$^{-1}$ Mpc$^{-1}$}

\def\msun{M$_{\odot}$}

\def\Hb{\mbox{${\rm H}{\beta}$}}
\def\Ha{\mbox{${\rm H}{\alpha}$}}
\def\OIIIa{\mbox{${\rm [O\,III]\,}{\lambda\,5007}$}}

\def\OII{\mbox{${\rm [O\,II]\,}{\lambda\,3727}$}}

\def\NII{\mbox{${\rm [N\,II]\,}{\lambda\,6584}$}}

\begin{document}

\title{The mass-metallicity relation of zCOSMOS galaxies at $z \approx 0.7$, its dependence on SFR, and the existence of massive low-metallicity galaxies\thanks{Based on observations
  obtained at the European Southern Observatory (ESO) Very Large
  Telescope (VLT), Paranal, Chile; ESO programs  084.B-0312A, 085.B-0317A and large program 175.A-0839}}


\author{C.~Maier\inst{1}
\and B.\,L.~Ziegler\inst{1}
\and S.\,J.~Lilly\inst{2}
\and T.~Contini\inst{3,4}
\and E.~P\'{e}rez-Montero\inst{5}
\and F.~Lamareille\inst{3,4}
\and M.~Bolzonella\inst{6}
\and E.~Le Floc'h\inst{7}
}

\institute{University of Vienna, Department of Astrophysics, Tuerkenschanzstrasse 17, 1180 Vienna, Austria\\
\email{christian.maier@univie.ac.at}
\and Institute of Astronomy, ETH Zurich, 8093 Zurich, Switzerland
\and Institut de Recherche en Astrophysique et Plan\'{e}tologie, CNRS, 14 avenue \'{E}douard Belin, 31400 Toulouse, France
\and IRAP, Universit\'{e} de Toulouse, UPS-OMP, Toulouse, France
\and Instituto de Astrof\'{i}sica de Andalucía, CSIC, Apartado de correos 3004, 18080 Granada, Spain
\and INAF – Osservatorio Astronomico di Bologna, via Ranzani 1, 40127 Bologna, Italy
\and CEA-Saclay, Orme des Merisiers, Bat. 709, 91191 Gif-sur-Yvette, France
}

\titlerunning{FMR at $z \sim 0.7$}
\authorrunning{C. Maier et al.}


\date{Received ; accepted}

\abstract 
{}{The knowledge of the number and of the physical nature of  low-metallicity massive galaxies is crucial for the determination and interpretation of the mass-metallicity relation (MZR).} 
{Using VLT-ISAAC near-infrared (NIR) spectroscopy of 39 zCOSMOS $0.5<z<0.9$ galaxies, we have measured H$\alpha $ and \NII\, emission line fluxes for galaxies with \OII, \Hb\, and \OIIIa\, 
available from VIMOS optical spectroscopy.
The NIR spectroscopy enables us to break the degeneracy of the  \R23\,  method to derive unambiguously O/H gas metallicities, and also star formation rates (SFRs) from extinction corrected \Ha\, fluxes.}
{
Using, as a benchmark, the position in the $D_{4000}$ vs. \OIIIa/\Hb\, diagram of galaxies with reliable O/Hs from NIR spectroscopy, we were able to break the lower/upper branch \R23\, degeneracy of additional 900 zCOSMOS galaxies at $0.5<z<0.9$, which do not have measurements of \Ha\, and \NII. 
Additionally, the \Ha-based SFR measurements were used to find the best SFR calibration based on \OII\, for the $0.5<z<0.9$ zCOSMOS galaxies without  \Ha\, measurements.
With a larger zCOSMOS sample of star-forming galaxies at $z \approx 0.7$ with reliable O/H and SFR measurements, we study
the MZR  at  $z \approx 0.7$  and its dependence on (specific) SFR.
We find a fraction of 19\% of lower mass $9.5<log(\rm{M}/\rm{M}_{\sun})<10.3$ zCOSMOS galaxies  which shows a larger evolution of the MZR relation, compared to higher mass galaxies, being more metal poor at a given mass by a factor of $2-3$ compared to SDSS. This indicates that the low-mass MZR slope is getting steeper at $z \approx 0.7$ compared to local galaxies.  
Many of these galaxies with  lower metallicity 
would be missed by studies which assume upper branch \R23\, solution for all galaxies. 
The existence of these metal-poor galaxies at $z \approx 0.7$
can be interpreted as the chemical version of galaxy downsizing.
Moreover, the sample of zCOSMOS galaxies shows direct evidence that the SFR influences
the MZR at these redshifts.
The comparison of the measured metallicities for the zCOSMOS sample with the values expected for a non-evolving fundamental metallicity relation (FMR) shows broadly agreement, and reveals that also galaxies with lower metallicities and typically higher (specific) SFRs, as found in our zCOSMOS sample at $z \approx 0.7$, are in agreement with the predictions of a non-evolving Z(M,SFR).}{}



\maketitle


\section{Introduction}

Metallicity is  one of the most fundamental properties of
galaxies. It is a measure of a galaxy's evolution, reflecting
the cycling of gas through stars, and any exchange of gas (inflows/outflows) between the galaxy and
its environment.  
Until inflows and outflows can be observed directly
and their mass flow rates can be quantified,
 measurements
of gas-phase metallicities and their relation to stellar mass can provide indirect insights into the impact of these gas flows on the chemical evolution of galaxies over time.


In the local universe, there is a tight 
mass-metallicity relation MZR \citep{lequeux79} 
with, in SDSS, a relatively steep slope below $M\sim 10^{10.7}$\msun, for a \citet{salp55} IMF, and flattening at higher masses \citep{trem04}.
Measurements of the 
metallicities 
of stars and star-forming gas in distant galaxies at
significant look-back times \citep[e.g.][]{hammer01,lilly03,kob03,maier04,maier05,maier06,maier14,lamareille09,maier14b,perez09b,zahid11} have mostly focussed
on the O/H abundance of star-forming gas as estimated from a number of empirically
calibrated metallicity estimators based on the relative strengths of strong emission
lines. These studies showed that the MZR at higher redshifts evolves relative to that seen locally. 
However, the observed shape and evolution of the MZR  is likely to be affected by the selection of the samples and metallicity estimators used.


To estimate the chemical abundances, a number of 
diagnostics have been developed based on strong lines, including the 
 $\rm R_{23}$ method
first introduced by \citet{pagel79}. This method is
based on the ratio of  ${\rm [O\,II]\,}{\lambda\,3727}$  and ${\rm [O\,III]\,}{\lambda\,5007}$ 
to H$\beta$, corrected for reddening and with the high/low
O/H degeneracy broken by the [NII]${\lambda\,6584}$/H$\alpha$ ratio.
\citet[][Ma05 in the following]{maier05} extended
the  $R_{23}$ method by simultaneously fitting all of these 5 lines ([OII], H$\beta$, [OIII], H$\alpha$ and [NII]) to derive 
metallicity O/H, reddening $A_V$ and ionization parameter $q$.  
%


Ma05 previously used VLT-ISAAC and Keck-NIRSPEC near-infrared (NIR) spectroscopy 
to measure ${\rm  H}{\alpha}$ and ${\rm [N\,II]\,}{\lambda\,6584}$ emission line (EL) fluxes  for
thirty $0.5<z<0.9$ galaxies extracted from the \emph{Canada France Redshift Survey} \citep{lilly95}.
Using the additional measured [OII], \Hb\, and [OIII] EL fluxes from optical spectra, Ma05 could measure
individual extinction values and reliable metallicities for these CFRS galaxies.
They found evidence for a population of galaxies with much lower metallicities than local galaxies with similar luminosities.
Their nature is still unclear, and
understanding why these probably massive galaxies exhibit low gas
metallicities is likely to give us insights into the chemical evolution
of the general galaxy population.
However, 
important additional information, 
namely the morphology and reliable stellar masses, was missing for the studied CFRS sample of galaxies (actually morphology
information was available just for one of the low-metallicity CFRS  galaxies).
This information is available for zCOSMOS galaxies  with possible low
metallicities and similar luminosities as the studied CFRS
galaxies: the morphology from ACS-HST images of the COSMOS field, and reliable stellar masses from a data set spanning a wide wavelength range extending to NIR.
Nevertheless, to establish the low metallicity of  zCOSMOS
objects, NIR spectroscopy is crucial to distinguish
between the upper branch  and  lower branch of the $R_{23}$ relation.
%


  \citet{perez13} used the calibration of the N2 parameter proposed by \citet{perez09} at $z<0.5$, and the conversion of the \R23\, relation  to this N2 parameter at $0.5<z<0.9$ to study the MZR of zCOSMOS galaxies. However, while zCOSMOS measurements of \OII, \Hb, and \OIIIa\, fluxes  from optical VIMOS spectra of $0.5<z<0.9$ galaxies were available, the \Ha\, and [NII] measurements required to break  the $R_{23}$ degeneracy were missing, so all objects were assumed to lie on the upper, high metallicity, \R23\, branch.
Moreover, no extinction correction 
using the Balmer decrement between \Ha\, and \Hb\, 
could be obtained for the $0.5<z<0.9$ sample studied by \citet{perez13}, and other less accurate extinction indicators had to be used. 


  In the local universe, the O/H at a given mass also depends on the SFR (and thus also SSFR) of the galaxy,
i.e., the SFR is a ``second-parameter'' in the MZR \citep[e.g.,][M10 in the following]{mannu10}. M10 also claimed that the Z(M,SFR) relation seen in SDSS is also applicable to higher redshift, calling this epoch-independent Z(M,SFR) relation the ``fundamental metallicity relation'' (FMR).
We refer the reader to \citet{maier14} for a discussion of the empirical formulations of the Z(M,SFR) by M10, of the physically-motivated formulation of the the Z(M,SFR) by \citet[][Li13 in the following]{lilly13}, and of the different extrapolations of these Z(M,SFR) relations to higher redshift.

\citet{perez13} took into account the dependence of metallicity on SFR to derive a SFR-corrected MZR for the zCOSMOS-bright sample. After this corection, an evolution of the SFR-corrected MZR could be still  observed, implying that  the FMR evolves with redshift.
On the other hand, \citet{cresci12} used a subsample of the 10k zCOSMOS sample \citep{lilly09} with a strong signal-to-noise (S/N) selection  
 and concluded no evolution of the FMR up to $z \sim 1$. 
In this paper, we try  to reconcile  these apparently contradictory results.



  The improvement in this paper compared to the \citet{perez13} and  \citet{cresci12} studies is due to  the new VLT-ISAAC NIR spectroscopy of \Ha\, and \NII\,  which enables us to derive reliable metallicities based on five ELs for a sample of 39 $0.5<z<0.9$ galaxies, and to use them as a \emph{benchmark to break the lower/upper branch $R_{23}$ degeneracy} of other zCOSMOS galaxies at $0.5<z<0.9$.  
Additionally, the SFRs based on the extinction corrected \Ha\, EL flux 
 in the sample with NIR spectroscopy can be also used as a \emph{benchmark to find the best SFR derivation method based on \OII\,} for the rest of the sample of $0.5<z<0.9$ galaxies, which does not have measurements of \Ha\, and \NII. Moreover, we can also  use the morphological information available for the zCOSMOS sample. 


  The paper is structured as follows: In Sect. 2 we
present the selection of a parent sample ELOX of zCOSMOS star-forming $0.5<z<0.9$ galaxies with ELs. The ISAAC-VLT observations of 39 galaxies selected from the sample ELOX are presented next.
In Sect.\,3 we investigate the AGN contribution
and present the derivation of SFRs, metallicities, extinction parameters, and stellar masses of the 39 galaxies with NIR spectroscopy.
 We also describe the calibration of the
SFRs of the larger ELOX sample, using the ISAAC data. 
In Sect.\,4 we first show how we can break the degeneracy of the \R23\, relation to derive metallicities of the ELOX sample using the NIR spectroscopy of the 39 galaxies. 
We then  present  the
SSFR-mass relation and MZR at $z\sim 0.7$, and investigate their dependence on 
morphologies. We also investigate if the SFR impacts the MZR, and how this compares with  predictions of the FMR from different studies.  
Finally in Sect.\,5 we present our conclusions.
A {\sl concordance}-cosmology with $\rm{H}_{0}=70$ \kmsmpc,
$\Omega_{0}=0.25$, $\Omega_{\Lambda}=0.75$ is used throughout this
paper.  
Note that {\sl
  metallicity} and {\sl abundance} will be taken to denote {\it oxygen abundance}, O/H, throughout this paper, unless otherwise specified.


\section{The data}

\subsection{The parent zCOSMOS sample}

The zCOSMOS project \citep{lilly07} has been securing spectroscopic 
redshifts for large numbers of galaxies in the
COSMOS field \citep{scoville07a}. 
The spectra of about 20\,500 $I_{AB} \le 22.5$ 
selected objects  over 
1.5 deg$^{2}$ of the COSMOS field, the so-called \emph{zCOSMOS-bright 20k sample}, are now all reduced.
 This sample includes about 1000\,stars, about 1500\,objects with no redshift determined,  and  about 18\,000\,galaxies at $0<z<1.4$ with redshifts measured with different confidence classes (reliabilities).

 The success rate of measuring redshifts
varies with redshift and is very high (more than 90\%) between $0.5 < z < 0.9$, the redshift range on which we concentrate  in this paper.   
Spectroscopic observations in zCOSMOS-bright were acquired using
VIMOS with the $\rm{R}\sim 600$ MR grism over the spectral range  5550-9650\AA. This enables
us to measure the \OII, \Hb, and \OIIIa\, line fluxes for galaxies at $0.5<z<0.9$.
For more details about the zCOSMOS survey, we refer the reader to \citet{lilly07} for the entire zCOSMOS survey, and to \citet{lilly09}, where the 
so-called \emph{zCOSMOS-bright 10k sample} is described, a zCOSMOS-bright sub-sample of about 10\,000 objects.


\subsection{Selection of EL zCOSMOS galaxies for metallicity studies}
\label{ELs}

From the parent 20k zCOSMOS sample we first exclude about 1000 stars, and then the broad line active galactic nuclei (AGNs) based on broad ELs seen in the VIMOS spectra.
Then, only galaxies with reliable redshifts are selected.
 Specifically, and with reference to the scheme described by \citet{lilly09}, we use reliable redshifts  with confidence classes 2.5, 3.x or 4.x,
 thus remaining with about 13500 galaxies at $0<z<1.4$. 
For the metallicities study we need  the \OII, \Hb, and \OIIIa\ lines to be covered by the VIMOS spectra; therefore we restrict the zCOSMOS sample to galaxies at redshifts $0.5<z<0.9$, leaving us with a sample of about 7000 objects.
About 200 X-ray  AGNs were then excluded using the XMM \citep{brusa07}, and Chandra COSMOS
observations \citep{elvis09}. 

Considering only objects with a S/N higher than two in the \OII, \Hb, and \OIIIa\,  EL fluxes, following the selection of \citet{perez13}, we remain with a sample of about 2150 EL galaxies.
We also exclude about 150  narrow line AGNs using the blue diagnostic diagram based on [OII], H$\beta$, and [OIII] and  Eq.1 from \citet{lam10}, as described in   \citet{perez13}, their Sect 3.1 and their Fig.\,1.
Thus the sample reduces to about 2000 galaxies. 
This is comparable to the number of galaxies used by \citet{perez13} in this redshift range (cf. their Table\,1).

For the O/H determinations we use a sub-sample of these  2000 galaxies, considering only flag 3.x and 4.x galaxies,
and requiring additionally that the [OII], \Hb\, and [OIII] lines are not  affected by strong sky lines \citep[cf. Sect.\,2.1.1 in][]{maier09}.
We also impose a stricter S/N selection threshold  in the EL fluxes  of  $S/N>3$ for [OIII] and  $S/N>5$ for H$\beta$, because \Hb\, and [OIII] are located in a region of the VIMOS spectrum affected by fringing.
The $S/N>3$ threshold for [OIII] is lower than for H$\beta$ in order not to bias the sample against high metallicity objects \citep[see Fig.\,4 in][]{foster12}.
 Exploring in detail many VIMOS spectra we found that these values for the S/N of the [OIII] and H$\beta$ fluxes 
minimize  the number of erroneous flux measurements and resulting wrong metallicities.
We chose a  $S/N>5$ for  H$\beta$ because 
closer inspection of spectra of zCOSMOS galaxies with $2<S/N<5$  in \Hb\, has revealed that most of them  have problematic  \Hb\,  in a noisy region of the VIMOS spectrum.
Including objects with $S/N<5$ in \Hb\, would imply that unreliable flux measurements with underestimated error bars (because of the fringing) would be used, which would produce spuriously low O/Hs. 
The applied $S/N>5$ selection threshold for  H$\beta$ thus minimises the contamination by spurious metallicities.
We thus remain with 939 galaxies in the \emph{sample ELOX} used for our O/H studies.
It should be noted that, after applying the cut of $S/N>5$ for \Hb, it turns out that 934 out of 939 galaxies, i.e., 99.5\% of the objects in the ELOX sample, have a $S/N>5$ for [OII].
As shown later in Sect.\,\ref{SSFRMsel}, where we discuss selection effects, our ELOX sample is slightly biased towards higher SSFR at lower masses, especially in the $0.62<z<0.75$ redshift bin; however, this selection bias applies also to other samples which select emission line galaxies to study metallicities, like the \citet{zahid11} sample we use for comparison.

\subsection{Sample selection for the ISAAC observations}
\label{sampleISAAC}

  We selected the ISAAC target galaxies from the \emph{sample ELOX}. 
For the selection of ISAAC targets we compared the expected observed wavelength of H$\alpha$
and [NII]$\lambda 6584$  ELs, accurately determined given the
zCOSMOS-bright velocity accuracy 
of $\rm \sim 100km s^{-1}$,
with the atlas of OH lines by Lidman et
al. from ESO. 
 We excluded objects where the H$\alpha$ or [NII]$\lambda 6584$ were expected to fall on strong  night sky lines.
Applying an additional H$\beta$ flux lower limit cut of $  4 \times 10^{-17}$ergs\,s$^{-1}$cm$^{-2}$ to restrict
  ISAAC integration times to less than one hour per object, we selected  39
  galaxies at
$0.5<z<0.9$ which we observed with ISAAC.


It turns out that 35 out of the 39 zCOSMOS galaxies
 are
at $0.5<z<0.75$ (see Table\,\ref{ISAACfluxes}).
Moreover, 797 ELOX galaxies (85\%  of the sample \emph{ELOX})
have redshifts $0.5<z<0.75$. 
Therefore we will concentrate our discussion on galaxies at $0.5<z<0.75$, and will sometimes split the sample in two redshifts bins to study also evolution and selection effects: 410 galaxies at $0.5<z<0.62$, and 387 galaxies at $0.62<z<0.75$.


\subsection{ISAAC observations and data reduction}
\label{obsdataredu}

NIR long-slit spectroscopy of 39  zCOSMOS galaxies at $0.5<z<0.9$ was
obtained with the ISAAC spectrograph at the VLT with the aim of measuring their
\Ha\,  and \NII\, EL fluxes.
The observations were carried out in February/March 2010 (Program 084.B-0312,
hereafter P84), and in April/May 2010 and January/February 2011
(Program 085.B-0317, hereafter P85). 
The medium resolution grism was used
with the Short-Wavelength channel equipped with a $1024 \times 1024$ pixel
Hawaii array. The pixel scale is 0.147\arcsec\, per pixel.
We used a slit of 2\arcsec\, width for the
P84 observations
 (seeing up to 1\arcsec.5 allowed, ESO priority B), and 1\arcsec\,
width for the
P85 observations (seeing up to 1\arcsec\,
allowed, ESO priority A), which results in a nominal resolution of
$\rm{R} \sim 1500$ for P84, and  $\rm{R} \sim 3000$ for P85, respectively.

We used two different filters, i.e., the SZ and the J filter, in order
to select the 5th and 4th grating order for measuring the \Ha\, and
\NII\, lines of galaxies at $0.5<z<0.65$ and $0.65<z<0.9$, respectively.
The covered wavelength range was 59\,nm and 46\,nm when using the J and the SZ
filter, respectively.  The corresponding pixel scales were $0.58$\AA/pixel and
$0.45$\AA/pixel.

The individual integration times varied between 1800\,s and 3600\,s,
depending on the brightness of the H$\beta$ EL flux, which
we used to estimate the minimum expected H$\alpha$ flux.
  During
the observations the telescope was nodded between two positions, A and B,
$\sim10$\arcsec\, apart along the slit.
 Dark frames, flat-fields and (Xe+Ar) arc lamp
spectra were taken with the same filter, central wavelength and slit width for
each of the targets observed during the night.  The conditions were
clear during these observations.

Spectra were reduced using the ESO ISAAC
pipeline, after removal of bias and of the cosmic rays in every
single exposure using the IRAF routine \emph{cosmicrays}.
The sequence with A and B representing the two positions along the slit was transformed in
ABBA by averaging the As and Bs. The pairs were then subtracted to give A-B, B-A, B-A and A-B. 
Wavelength calibration was performed using OH skylines.
The subtracted frames contain positive and negative spectra. The two spectra were combined and the resulting
frames were then added together to give the final result.
Flux calibration was done using the broadband magnitudes
of the telluric standards.
The one-dimensional spectra of each galaxy were extracted 
 using the algorithm by \citet{horne} with an aperture of 10-15 pixels in y-direction, i.e. about
1.5$-$2\arcsec.  
For more details on data reduction of ISAAC observations we refer the reader to Sect.\,2 in  Ma05, where we dealt with very similar VLT ISAAC spectroscopy of CFRS galaxies at similar redshifts
as the zCOSMOS-bright target galaxies.


\section{The Measurements}


\subsection{EL Fluxes}
\label{emline}

  EL fluxes were measured from the calibrated
NIR ISAAC spectra
following the procedure described in Ma05. 
Table \ref{ISAACfluxes} reports, for each individual galaxy, the EL fluxes (or upper limits).
Specifically:

  {\it (i)}  We were able to measure \Ha\, in the NIR for all 39 galaxies
reported in Table\,\ref{ISAACfluxes}. We neglected Balmer absorption 
in estimating the intrinsic \Ha\, fluxes. 
For a wide range of star formation histories and for plausible IMFs, the stellar absorption at \Ha\, is always $<$5\AA\, \citep[see, e.g.][]{brinchmann04}, small compared to the equivalent widths of the feature in emission in our  galaxies and to other sources of uncertainties.

  {\it (ii)} [OII], \Hb, and [OIII] fluxes were measured for all 39 galaxies using
the automatic routine Platefit\_VIMOS (Lamareille et al. 2015, in
preparation).
After removing a stellar component using \citet{bruzcharl03}
models, Platefit\_VIMOS performs a simultaneous fit of all
ELs using a Gaussian profile.
The  \Hb\, flux was thereby corrected for Balmer absorption, with typical values between 1 and 7\AA\, used to correct the equivalent width of \Hb. 
Additionally, an aperture correction to take into account slit losses was applied to each VIMOS spectrum, as described in Sect.\,2.1.6 of \citet{maier09}.
Each zCOSMOS spectrum was thereby convolved with the ACS I(814) filter and then
this magnitude was compared with the measured I-band magnitude of the respective galaxy. 
The difference between the two magnitudes gives the aperture
correction factor for each spectrum, with typical aperture correction factors between 1 and 3.
This correction assumes that the respective emission line flux and I-band continuum suffer equal slit losses.

  {\it (iii)} The \NII\, EL flux was measured for 18 galaxies, but was too faint to be detected for 21
objects. Nevertheless, we
could determine upper limits for the \NII\, EL flux of these 21
galaxies, because the ISAAC targets were selected such that the [NII] line should avoid strong night sky lines (as described in Sect.\,\ref{sampleISAAC}).
 Table\,\ref{ISAACfluxes} lists the
$2\sigma$ upper limits for these 21 galaxies.


\begin{table*}
\caption{Observed and derived quantities for the 
 zCOSMOS $0.5< z < 0.9$ galaxies with NIR follow-up spectroscopy}\label{ISAACfluxes}
\begin{tabular}{cccccccccc}
\hline\hline      
Id & z &   [OII]\tablefootmark{a}    &    \Hb\tablefootmark{a}
  &   [OIII]\tablefootmark{a}    &   \Ha\tablefootmark{a}    &  [NII]\tablefootmark{a,b}     &log(M/M$_{\sun}$)& 
  O/H &     A$_{\rm V}$\\
\hline      
803717&0.6976&  23.51 $\pm$  0.37 & 12.20 $\pm$  0.69 & 11.65 $\pm$  0.84 & 57.2 $\pm$  5.2 & 13.0 $\pm$  5.2  &10.71$^{+0.05}_{-0.06}$  &8.96$^{+0.02}_{-0.02}$&1.24$^{+0.29}_{-0.29}$\\
803924&0.7927&  17.97 $\pm$  0.35 &  4.86 $\pm$  0.43 & 20.92 $\pm$  1.85 & 30.0 $\pm$  2.0 &  $<$1.8          &10.09$^{+0.08}_{-0.03}$  &8.50$^{+0.05}_{-0.06}$&1.01$^{+0.15}_{-0.18}$\\
804474&0.7137&  13.04 $\pm$  0.35 &  6.46 $\pm$  0.90 & 12.56 $\pm$  0.66 & 23.4 $\pm$  2.6 &  $<$3.9          &10.21$^{+0.08}_{-0.09}$  &8.35$^{+0.20}_{-0.28}$&0.73$^{+0.54}_{-0.51}$\\
\hline
\end{tabular}
\tablefoottext{a}{Fluxes are given in $10^{-17}\rm{ergs}\,\rm{s}^{-1}\rm{cm}^{-2}$.}\\
\tablefoottext{b}{ $2\sigma$ upper limits for the EL flux of \NII\, are given if [NII] is not detected.}\\
\tablefoottext{c}{Double-valued O/H solution.}\\
(The entire version of this table for the full sample of 39 galaxies will be available after the paper is published in A\&A.)
\end{table*}


Given the UV to IR coverage of the SED of the zCOSMOS objects with 30
bands \citep{ilbert09}, 
we are able to reliably compute the NIR  
and optical magnitudes from the SED to perform the relative calibration of the
EL fluxes measured in the optical and NIR, also in the
case of non-photometric spectroscopic observations.
Similar to  point (ii) above, we also convolved the ISAAC zCOSMOS spectra (we detect the continuum for most ISAAC targets) with the respective NIR filter and compared the resulting magnitude with the measured NIR magnitude, to obtain the aperture correction for the ISAAC spectra.
In a few cases of the P84 observations we do not detect the continuum in NIR; however, since we used a slit of 2\arcsec\, for these observations, the resulting EL fluxes should approximate ``total'' fluxes (cf. discussion in Sect.\,2 of Ma05). 
The reliability  of the relative calibration between the NIR and optical data (necessary to measure the extinction from the Balmer decrement) is reinforced by the good agreement  between the SFRs derived from extinction corrected H$\alpha$ and the SFRs derived from IRAC and MIPS infrared data (Sect.\,\ref{SFRs}).


\subsection{The BPT diagram}

\begin{figure}
\includegraphics[width=8cm,angle=270,clip=true]{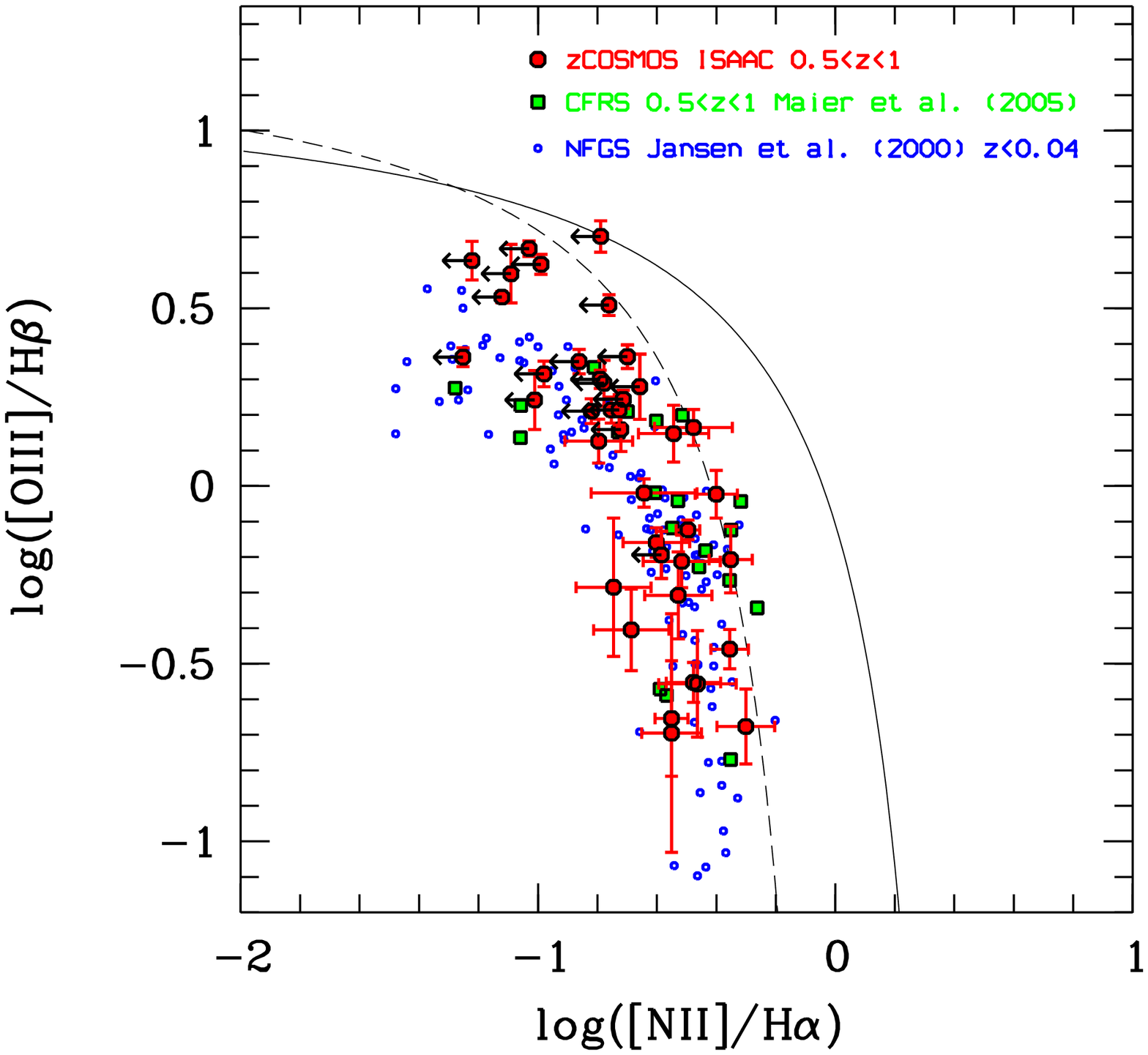}
\caption
{
\label{BPT_ISAAC} 
\footnotesize 
BPT \citep{bald81} diagnostic diagram to disentangle star formation-dominated galaxies
from AGNs. The zCOSMOS galaxies  with ISAAC spectroscopy are shown as filled red circles, $z<1$ CFRS galaxies from Ma05 as filled green squares, and local
NFGS galaxies from \citet{jansen} as blue points.
The zCOSMOS $0.5<z<0.9$ galaxies
lie under and to the left of the theoretical (solid) curve of \citet{kewley01} and of the empirical (dashed) curve of \citet{kauf03},
 which separate star-forming galaxies (below/left of the
curves) from AGNs (above/right of the curves).
}
\end{figure}

While \citet{perez13} had to use the ``blue BPT'' \citep{lam10} diagram [OIII]/\Hb\, vs. [OII]/\Hb\, to identify Type-2 AGNs at $0.5<z<0.9$, the NIR spectroscopy enables us to use the original BPT \citep{bald81} diagram [OIII]/\Hb\, vs. [NII]/\Ha\, shown in Fig.\,\ref{BPT_ISAAC},
to establish if the source of gas ionization is of stellar origin,
or rather associated with AGN activity.
The  $0.5<z<0.9$ zCOSMOS galaxies with ISAAC spectroscopy
lie under and to the left of the theoretical (solid) curve of \citet{kewley01}, which separates star-forming galaxies from AGNs, and they lie also under and to the left of the empirical (dashed) curve of \citet{kauf03}. 
This indicates that in  our $0.5<z<0.9$ zCOSMOS galaxies with ISAAC spectroscopy
the dominant source of ionization in the gas is recent star
formation.

As described in  Sect.\,\ref{ELs}, we excluded narrow line AGNs from the ELOX parent sample using the ``blue BPT diagram''  from \citet{lam10}.
Therefore, the findings of Fig.\,\ref{BPT_ISAAC} indicate that the ``blue BPT diagram'', as used by \citet{perez13}, is efficient in removing type-2 AGNs in the zCOSMOS sample.

\subsection{Derivation of Oxygen Abundances and Extinction Parameters}
\label{Oxabund}

 To derive oxygen abundances O/H
and the extinction parameters $A_{V}$ for the 39  zCOSMOS galaxies with
NIR ISAAC spectroscopy we used the approach described in Ma05.
This is based on the models of \citet[][KD02]{kewdop02}, who developed a set
of ionisation parameter and oxygen abundance diagnostics based on the use of
strong rest-frame optical ELs.  The method relies on a
simultaneous fit to all available EL fluxes (including the
\NII\, upper limits) in terms of extinction
parameter $A_{V}$, ionisation parameter $q$, and O/H, and is 
described in detail in Ma05. It should be noted that, as described in Sect.\,3.1 of Ma05, we use the wavelength-dependent reddening
function  $f(\lambda)$ of \citet{whitford}.

\citet{maier05,maier06} showed that, if [OII], H$\beta$, [OIII] and H$\alpha$ are all measured, an upper limit for [NII]$\lambda 6584$ often suffices to break the O/H degeneracy and determine oxygen abundances based on the $\chi^{2}$ analysis presented in Ma05.
The O/H degeneracy is due to the fact that $R_{23}$ is not a monotonic function of oxygen abundances, i.e., it is double-valued: at a given value of $R_{23}$ there are two possible choices of the oxygen abundance.
The point of the line-fitting approach developed in
Ma05 is to use all of the data and the best available emission line models (in this case KD02) without applying other priors that may or may not be valid, in order to compute O/H and thereby also breaking the $R_{23}$ degeneracy if the data quality allows.
The resulting metallicities for the zCOSMOS galaxies with NIR spectroscopy are given in Table\,\ref{ISAACfluxes}, including six out of 39 galaxies with a double-valued O/H measurement.
The data quality did not suffice, i.e., the upper limits of [NII]$\lambda 6584$ of these six galaxies were not enough to break the upper/lower branch degeneracy.

The error bars of O/H and $A_{V}$ that we report in  Table\,\ref{ISAACfluxes}
are the formal 1$\sigma$ confidence intervals for the projected best-fitting values.
As described in Sect.\,3.3 of Ma05, the error bars of the oxygen abundance  (or $A_{V}$) are
given by  the range of oxygen abundance (or  $A_{V}$, respectively) for those 
models with $ \chi^{2}$ in the range  $\chi_{min}^{2} \leq \chi^{2} \leq
\chi_{min}^{2} +1 $ (where  $\chi_{min}^{2}$  is the minimum $\chi^{2}$
of all allowed models for a given galaxy), corresponding to  a confidence level of 68.3\%
for one single parameter. 
A detailed discussion of the calculation of the error bars of O/H and $A_{V}$,
and of the assignment of non-degenerate (not double-valued) oxygen abundances is given by Ma05, and also in Sect.\,3.5 of  \citet{maier14}, so we refer the reader to these papers.

We would like to mention here that, in terms of absolute measurements of metallicity, there are two
sources of uncertainties: 
(i) The purely statistical measurement
uncertainties propagating through to the parameter determinations. These
are addressed by our $\chi^{2}$ analysis.  They reflect both the quality 
of the data and the gradients (and degeneracies) in the models.
These uncertainties are presented in  Table\,\ref{ISAACfluxes}. 
(ii) Uncertainties in the KD02 models. Application 
of many different \R23\, calibrations to the SDSS data \citep{keweli08} indicates 
a range of $\pm 0.2$\,dex in the mean O/H at a given mass -
with the KD02 models more or less in the middle. This latter problem is 
reduced by using the same analysis on all objects and focussing on
differential effects with $z$ or M, rather than estimates of the absolute
metallicity.
Our philosophy is to treat all galaxies at different redshifts and in different samples  in
the same way and to focus on relative effects between the selected samples. 
Therefore, when comparing metallicities of the different samples,  we think it is more appropriate to  
consider  the uncertainty (i) described above.


\subsection{Stellar Masses}

To perform SED fitting analysis and estimate the galaxy stellar mass based on the optical to IRAC photometric data from COSMOS, the \emph{HyperzMass} software was used.
This is a modified version of the public photometric redshift code \emph{Hyperz} \citep{bolzo00}.
\emph{HyperzMass} fits photometric data points with the \citet{bruzcharl03} synthetic stellar population models, and picks the best-fit parameters by minimizing the $\chi^{2}$ between observed and model fluxes. 
The stellar mass is obtained by integrating the SFR over the galaxy age, and correcting for the mass loss in the course of stellar evolution.

The choice of the different parameters was discussed and compared by \citet{bolzo10} for the $z<1$ redshift regime,  
 so we refer the reader to that paper for details.
Specifically, \citet{bolzo10} showed that the stellar mass is a rather stable parameter in SED fitting for $z<1$ galaxies when dealing with a data set like COSMOS, spanning a wide wavelength range extending to NIR.
It should be noted that  the stellar mass calculated in this paper is always  the actual mass of long-lived stars, and 
when literature data is used we always correct to this definition if necessary.


\subsection{Star formation rates (SFRs)}
\label{SFRs}
One of the most reliable and well calibrated SFR indicators is the
\Ha\, EL.
For the 39 zCOSMOS galaxies with NIR spectroscopy, the \Ha\, luminosities (corrected for extinction using the $A_{V}$ values computed as described in Sect.\,\ref{Oxabund})
were used to calculate the SFRs, by applying  the \citet{ken98} conversion of H$\alpha$ luminosity into
SFR: $\rm{SFR} (M_{\odot}\rm{yr}^{-1}) = 7.9 \times 10^{-42}
\rm{L}(\rm{H}\alpha)\rm{ergs/s}$.

\begin{figure*}
\centering
\includegraphics[width=15cm,angle=270,clip=true]{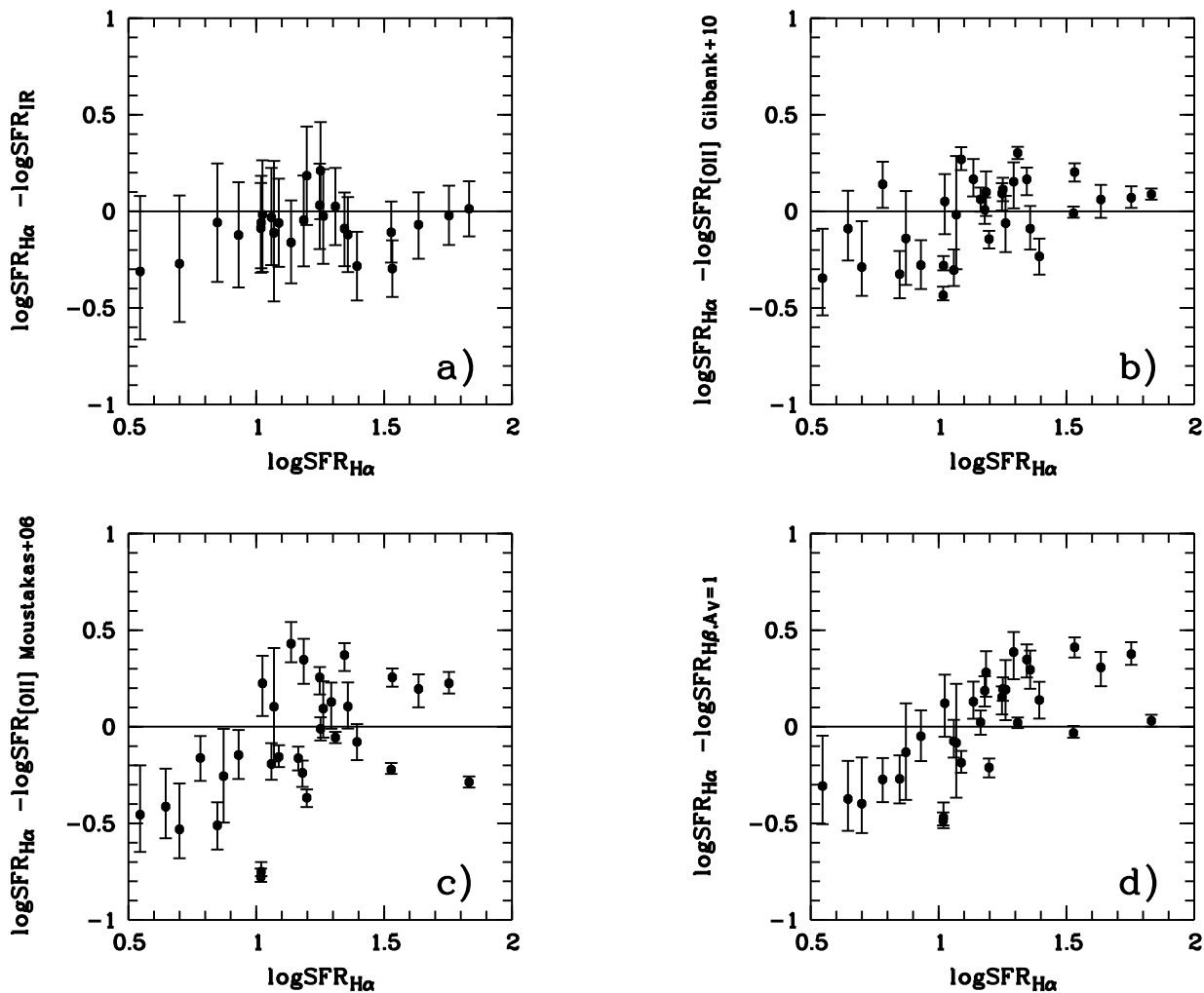}
\caption
{
\label{Vgl_SFROII} 
\footnotesize 
Comparison of SFRs computed using different methods for zCOSMOS galaxies
with ISAAC NIR spectroscopy.
Panel a  shows a good agreement (within the measurements errors) of the
SFRs from ISAAC H$\alpha$ and SFRs from  infrared COSMOS IRAC and MIPS data for the objects with infrared detections (Le Floch et al. 2015, in preparation).
The panels b-d  show the comparison of H$\alpha$-based SFRs with: 
b)  SFRs from the [OII] luminosity, applying the \citet{gilb10} calibration, their Eq.\,8, and taking into account the erratum \citet{gilb11} of that paper;
c) SFRs derived from the [OII] luminosity using equation (2) of \citet{maier09}, based on the study of \citet{moust06};
and d) SFRs from the H$\beta$ luminosity, assuming $A_{V}=1$ (an assumption often used in literature when the Balmer decrement cannot be measured)
to derive H$\alpha$ luminosities.
The  best agreement between the reliable, extinction corrected, \Ha\,-based SFRs and SFRs estimators based on  other ELs (panels b-d) is in panel b.
}
\end{figure*}

Fig.\,\ref{Vgl_SFROII} shows the comparison of the SFRs computed using different methods, for zCOSMOS galaxies
 with ISAAC NIR spectroscopy. 
SFRs from COSMOS IRAC and MIPS infrared data  for the objects with infrared detections were calculated converting the rest-frame 8$\mu m$ luminosity to a dust-corrected Paschen-$\alpha$ luminosity
using the calibration from \citet{calz05}, and then converting the  Paschen-$\alpha$ luminosity to SFR using a calibration from   \cite{osterb89} (for more details see Le Floch et al. 2015, in preparation).
The mean value of the difference between SFRs from  H$\alpha$ and SFRs from  COSMOS IRAC and MIPS is -0.075 dex (panel a of Fig.\,\ref{Vgl_SFROII}) indicating that SFRs from  H$\alpha$ are typically smaller than SFRs from COSMOS IRAC and MIPS by about 16\% . However, given the large error bars (see panel a of Fig.\,\ref{Vgl_SFROII}), the two different methods to measure SFRs  are still consistent with each other.
This is reassuring, and reinforces the reliability of the H$\alpha$-based SFRs and of the relative calibration between the NIR and optical data required to measure the extinction from the Balmer decrement (cf. Sect.\,\ref{emline}).

Most of the zCOSMOS-bright galaxies at $0.5<z<0.9$ do not have NIR measurements of \Ha, i.e., their SFRs have to be determined using bluer ELs observed with VIMOS, like \Hb\, or [OII]. 
We compare  the  \Ha\,-based SFRs with SFRs estimators using \Hb\, and [OII] in panels b-d of Fig.\,\ref{Vgl_SFROII}.
Panel b shows a slight trend towards increasing logSFR$_{\Ha}$-log(SFR)$_{[OII]}$ for higher logSFR$_{\Ha}$, but this trend is more pronounced in panels c an d (the value for the slope of a linear fit to the data is about two times higher in panels c and d compared to panel b). Moreover, taken into account the error bars, twice as many objects in panels c and d are more than 1$\sigma$ away from agreement (zero line) than in panel b.
 Therefore, in the following, when \Ha\, is not available for ELOX galaxies, we will use SFRs derived from  the [OII] luminosity using the \citet{gilb10} calibration (panel b). It should be noted that the \citet{gilb10} [OII]-SFR calibration compensates for the effects of metallicity dependence and dust extinction by applying an empirical mass-dependent correction for the [OII]-SFR relation, which was derived using data in the deeper SDSS Stripe 82 subsample. 


\subsection{Morphological classification}

We use the 
morphological classification
of the COSMOS galaxies based on the Zurich Estimator of
Structural Types (ZEST) classification derived by \citet{scarlata07}.
The ZEST classification scheme is based on the principal
component analysis of five non-parametric diagnostics, i.e., asymmetry, concentration, Gini coefficient, second-order moment of the brightest 20\% of galaxy pixels, and
ellipticity. The ZEST scheme morphologically classifies galaxies into
three main types: early type galaxies T1, disk galaxies T2, and irregular
galaxies T3.
It should be noted that the ELOX sample, selected to be a sample of galaxies with ELs,  contains only 13 galaxies ($<1.5$\% of the sample) classified as ZEST early-type T1; therefore we will concentrate on the types T2 and T3 when discussing morphologies of the ELOX sample.


\section{Results}


\subsection{Breaking the Degeneracy in $R_{23}$ for ELOX Sample Galaxies without NIR Spectroscopy}
\label{R23degen}

\begin{figure}
\includegraphics[width=8cm,angle=270,clip=true]{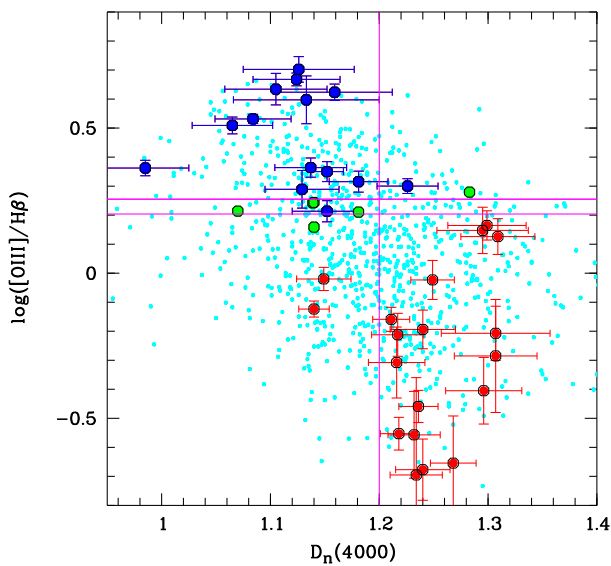}
\caption
{
\label{CaliOH} 
\footnotesize 
The $D_{4000}$ vs. \OIIIa/\Hb\, diagram used to break the \R23\, degeneracy for
the $0.5<z<0.9$ zCOSMOS galaxies (cyan points) without NIR spectroscopy.
The  39 zCOSMOS galaxies with ISAAC spectroscopy  are shown as blue filled circles for galaxies with low (branch) O/H measurements,  red filled circles for 
galaxies with high (branch) O/H measurements, and green filled circles for the six galaxies with double-valued solutions (note that the  $D_{4000}$ and \OIIIa/\Hb\, values of two galaxies are almost identical, i.e., the respective green circles coincide).
This diagram can be used to
break the lower/upper branch \R23\, degeneracy of other $0.5<z<0.9$  galaxies  without NIR spectroscopy, but similar $D_{4000}$  and [OIII]/\Hb\, as the 33 benchmark galaxies with ISAAC spectroscopy and reliable metallicities.
}
\end{figure}

To  break the \R23\, degeneracy 
for the 900 galaxies in the ELOX sample without NIR spectroscopy of \Ha\, and \NII, one needs an additional indicator. 
We use the $D_{4000}$ vs. \OIIIa/\Hb\, diagram for this purpose (Fig.\,\ref{CaliOH}).
%

The $D_{4000}$ vs. \OIIIa/\Hb\, values for the zCOSMOS ELOX sample at $0.5<z<0.9$
are shown as cyan points in Fig.\,\ref{CaliOH}.
The  33 zCOSMOS galaxies with reliable O/H measurements are shown as blue filled circles for galaxies with low (branch) O/H measurements, and red filled circles for 
galaxies with high (branch) O/H measurements. Additionally, green filled circles depict the six galaxies with double-valued solutions mentioned in Sect.\,\ref{Oxabund}.
There is a region of the $D_{4000}$ vs. \OIIIa/\Hb\, diagram with $D_{4000}<1.2$ and [OIII]/H$\beta >1.8$ (above the upper horizontal magenta line) occupied by galaxies with low (branch) O/Hs (blue symbols), and a region with $D_{4000}>1.2$ and [OIII]/H$\beta <1.6$ (below the lower horizontal magenta line) occupied by galaxies with high (branch) O/Hs (red symbols).
This information can be used to
break the lower/upper branch \R23\, degeneracy of other zCOSMOS ELOX  galaxies  without NIR spectroscopy, but similar $D_{4000}$  and [OIII]/\Hb\, as the 33 benchmark galaxies with ISAAC spectroscopy and reliable metallicities.
There are only a few objects in the upper right corner of Fig.\,\ref{CaliOH} ($D_{4000}>1.2$ and [OIII]/H$\beta >1.8$), $\sim 5$\% of the ELOX sample, and these are assumed to be on the upper \R23\, branch, to also account for the fact that a few galaxies with [OIII]/H$\beta >1.8$ are expected to be on the upper \R23\, branch according to Fig.\,\ref{BPT_OIIIHb}.
For the ELOX galaxies in the lower left corner of Fig.\,\ref{CaliOH} ($D_{4000}<1.2$ and [OIII]/H$\beta <1.8$) we assume that they are on the upper \R23\, branch, because of two indications: (i) the two galaxies in this corner with ISAAC measurements have upper \R23\, branch metallicities (red filled circles), and  (ii) Fig.\,\ref{BPT_OIIIHb} in appendix indicates that they should have $O/H>8.6$, because EL galaxies along the star-forming sequence (between the dashed lines) with [OIII]/H$\beta <1.8$ do not lie in the cyan colored grid area ($O/H<8.6$) in Fig.\,\ref{BPT_OIIIHb}.

Our empirical finding based on the zCOSMOS galaxies with ISAAC spectroscopy is that objects with $D_{4000}<1.2$ and [OIII]/H$\beta >1.8$ lie on the lower \R23\, branch, when  computing O/Hs based on the KD02 models and using 5 ELs. Our findings are based on the particular form of the KD02 models, especially the particular form of the \R23\, relation (Fig.\,5 in KD02). Using a different parametrization of the \R23\, relation may possibly give different results. Therefore, we advise caution when using Fig.\,\ref{CaliOH} to select the \R23\, branch: larger samples of galaxies with 5 ELs measured are needed to verify that the proposed diagram can really be used to select the \R23\, branch also for other samples and also for other metallicity calibrations.
We are working on this  and also exploring the reason why the  $D_{4000}$ vs. [OIII]/\Hb\, diagram is able to distinguish the \R23\, branch (and if it always works) using a larger sample of galaxies from the Clash-VLT survey, at slightly lower redshifts $0.4<z<0.5$. This sample has reliable metalllicities based on all five ELs measured with VIMOS at the same time, including [NII] flux measurements, and not only upper limits, also for high [OIII]/\Hb\, values.  This work will be presented in a future paper \citep[Maier et al. 2015, in preparation, first results shown in][]{kuchn14}.

It should be noted that this method to break the $R_{23}$ degeneracy for $\sim 10^{10}$\msun\, galaxies can not be derived/improved using the larger SDSS local sample of galaxies, because virtually no galaxy in the SDSS sample has such a high \OIIIa/\Hb\, line ratio \emph{at similar stellar masses} like the low metalllicity zCOSMOS galaxies with ISAAC spectroscopy shown in Fig.\,\ref{CaliOH}. This is indicated, e.g., by the orange contours in the upper panel of Fig.\,7 of \citet{holdoesch14}, depicting the \OIIIa/\Hb\, line ratios as a function of stellar mass for star-forming (not dominated by AGNs) SDSS galaxies.  

As described in Sect.\,\ref{sampleISAAC}, the galaxies with ISAAC spectroscopy were selected to have a minimum \Hb\, flux limt. Nevertheless, Fig.\,\ref{SSFRMassz07} shows that  ISAAC target galaxies  (big coloured squares and triangles) occupy a similar region of the SSFR-mass diagram as the ELOX sample. Therefore, there should be no selection bias of the ISAAC sample compared to the ELOX sample.

To get O/H for the ELOX galaxies without NIR spectropscopy of \Ha\, and [NII], but [OII], \Hb\, and [OIII] fluxes measured with VIMOS, we proceed as follows.
As described in Ma05,  [NII]/\Ha\, basically only determines the \R23\, branch, and the \R23\, diagnostic is used to determine the exact value of the metallicity on the respective branch. Therefore, we do not need the value of [NII]/\Ha, but we use instead Fig.\,\ref{CaliOH} to estimate which branch is appropriate for each galaxy, as described above.  When using our $\chi^2$ analysis presented in Sect.\,\ref{Oxabund},  we allow the $\Ha/\Hb$ ratio to vary a priori in a typical range $2.86-6$, corresponding to typical extinction values of star-forming galaxies of $0<A_{V}<2.1$, as found for the zCOSMOS galaxies with ISAAC spectroscopy (see Table\,\ref{ISAACfluxes}).  This way, we get an estimate for metallicity O/H, reddening $A_V$ and ionization parameter $q$ for the entire ELOX sample.
%

\subsection{The SSFR-mass relation at $z \approx 0.7$ and selection effects}
\label{SSFRMsel}

\begin{figure*}
\centering
\includegraphics[width=12cm,angle=270,clip=true]{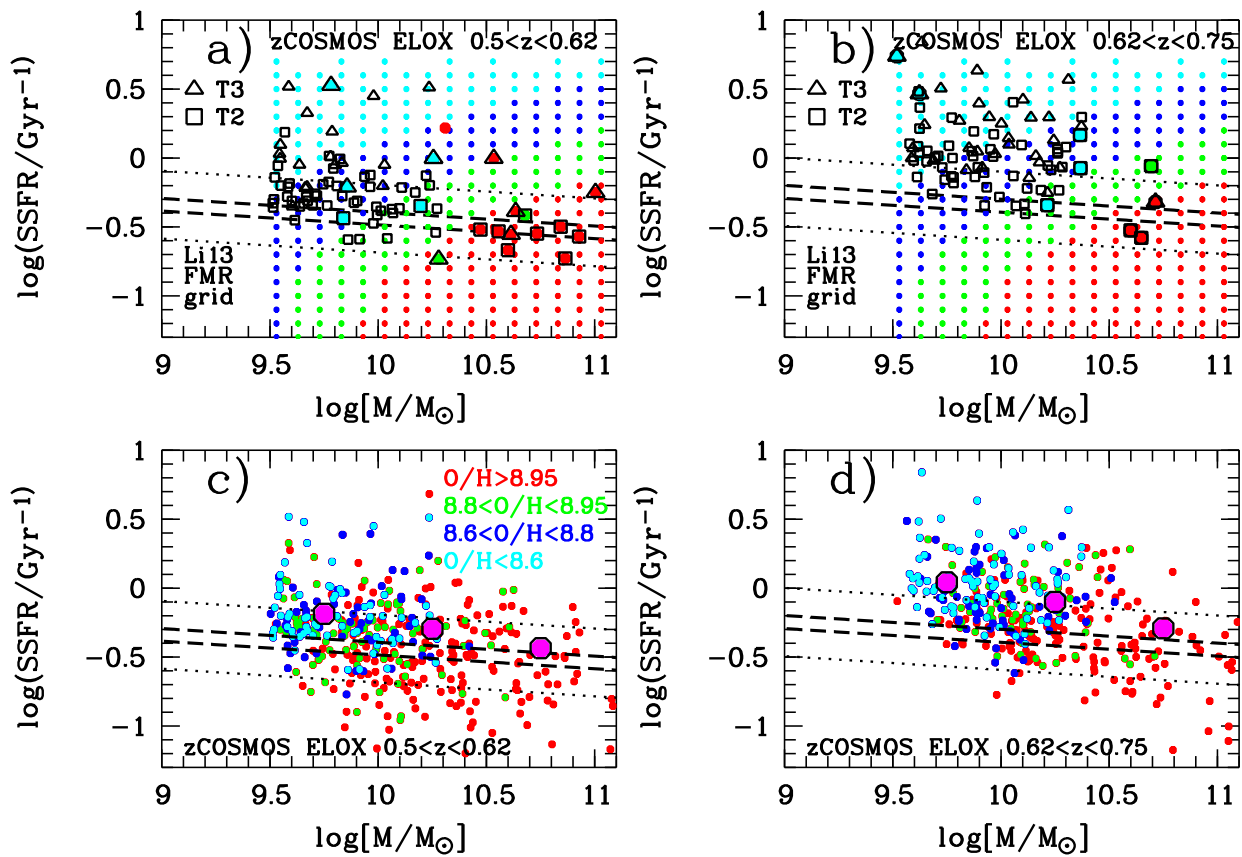}
\caption
{
\label{SSFRMassz07} 
\footnotesize 
The SSFR-mass relation for zCOSMOS galaxies at $0.5<z<0.62$ (panels a and c) and at $0.62<z<0.75$ (panels b and d), with the data points color-coded according to the measured metallicity value (legend in panel c).
The colors of the grid points in panel a and b indicate the metallicities expected for the Z(M,SFR) extrapolation of Li13.
In panels a and b, galaxies with ISAAC spectroscopy are shown as colored symbols, while ELOX galaxies with $O/H<8.6$ (the respective cyan circles in panels c and d) are shown as black symbols as squares (triangles) for T2 (T3) morphologies.
In panels c and d the big magenta filled circles show the mean values of the SSFRs in three mass bins.
ELOX galaxies with lower \R23\, branch KD02 metallicities $O/H<8.6$,
have typically lower masses ($9.5<log(\rm{M}/\rm{M}_{\sun})<10.3$), and preferentially irregular morphologies when they have higher SSFRs.
This diagram also indicates that SFR is a second parameter in the MZR of the zCOSMOS sample at $0.5<z<0.75$, because galaxies with higher (S)SFRs have generally also lower metallicities.
}
\end{figure*}

  The specific SFR (SSFR) has been found to be a tight but weak function of mass
at all epochs up to $z \sim 2$ for (most) star-forming galaxies
(\emph{main sequence, MS}), and to
evolve strongly with epoch by a factor of 20 to $z\sim2$ \citep{daddi07,elbaz07,noeske07,salim07,peng10}.
The dashed black (almost horizontal) lines in Fig.\,\ref{SSFRMassz07} show the SSFRs calculated using  Eq.1
of \citet{peng10} for the MS at $z=0.5$ and $z=0.62$ (left panels),
and at $z=0.62$ and $z=0.75$ (right panels), assuming a weak dependence of SSFR on mass (as observed for local SDSS galaxies), SSFR$\propto$M$^{\beta}$ with $\beta=-0.1$. The dispersion of $\sim 0.3$\,dex about the mean relation is taken into account by the dotted (almost horizontal) lines.
We  consider as  MS objects those galaxies 
in the region between the dotted lines in Fig.\,\ref{SSFRMassz07}.

SFRs for the  zCOSMOS galaxies with ISAAC NIR follow-up (large colored symbols in panels a and b in Fig.\,\ref{SSFRMassz07}) were derived from \Ha, while for the parent ELOX sample of 797 galaxies at $0.5<z<0.75$ (panels c and d) we used SFRs derived from [OII] as described in Sect.\,\ref{SFRs}.  
The big magenta filled circles in panels c and d in Fig.\,\ref{SSFRMassz07} indicate the mean SSFR in three mass bins $9.5\le log(M/M_{\odot}) \le 10$, $10<log(M/M_{\odot})\le 10.5$, and $10.5<log(M/M_{\odot})\le 11$.  The position of the mean SSFR relative to the MS in Fig.\,\ref{SSFRMassz07} indicate that both samples at $0.5<z<0.62$ and at $0.62<z<0.75$  are quite representative for MS galaxies at the higher masses.
 At lower masses,  the $0.5<z<0.62$ sample is still dominated by MS galaxies, while  $0.62<z<0.75$ galaxies lie predominantly  above the MS, which could bias the low mass $0.62<z<0.75$ sample more towards lower metallicities.


The data points in Fig.\,\ref{SSFRMassz07} are color-coded according to their measured metallicity values (legend in panel c).
The colors of the grid points in panel a and b indicate the metallicities derived for the Z(M,SFR) from Eq.\,40 in Li13.
This way one can explore if the measured metallicities (color of the \emph{data} points) and the expected metallicity assuming a redshift independent Z(M,SFR) relation (color of the \emph{grid} points) are in agreement.
Like in \cite{maier14}, because M10 and Li13 use a \citet{chabrier03} IMF, we converted the masses to a \citet{salp55} IMF  after computation of the expected O/Hs (color of the grid points), by adding 0.23\,dex to the stellar masses.

It should be noted that, when using the KD02 calibration, the upper metallicity \R23\, branch roughly corresponds to $\rm{O/H}>8.6$ (see Fig.\,5 in KD02).
The  ELOX low metallicity galaxies with $O/H<8.6$ (cyan circles in panels c and d) are shown again as black symbols in panels a and b, to explore their nature. 
Many of these galaxies with  lower metallicity 
would be missed by studies which assume upper branch \R23\, solution for all galaxies. 
They represent $\sim 19$\% of the ELOX sample and have typically lower masses ($9.5<log(\rm{M}/\rm{M}_{\sun})<10.3$), and preferentially irregular morphologies when they have higher SSFRs.

Our study shows that these low metallicities galaxies exist at intermediate redshifts because:
i) first, they lie in a region of the SSFR-mass diagram where also some zCOSMOS galaxies with ISAAC spectroscopy and \emph{reliable} $O/H<8.6$ are located; ii) second, given their masses and SFRs, the prediction of the Z(M,SFR) extrapolation of Li13 is $O/H<8.6$ for most of these galaxies, as shown by the cyan color of the grid points in panels a and b in the region occupied by the majority of the black symbols (which depict galaxies with  $O/H<8.6$).

\subsection{The MZR at $z \approx 0.7$}
\label{sec:MZR07}

\begin{figure*}
\includegraphics[width=8cm,angle=270,clip=true]{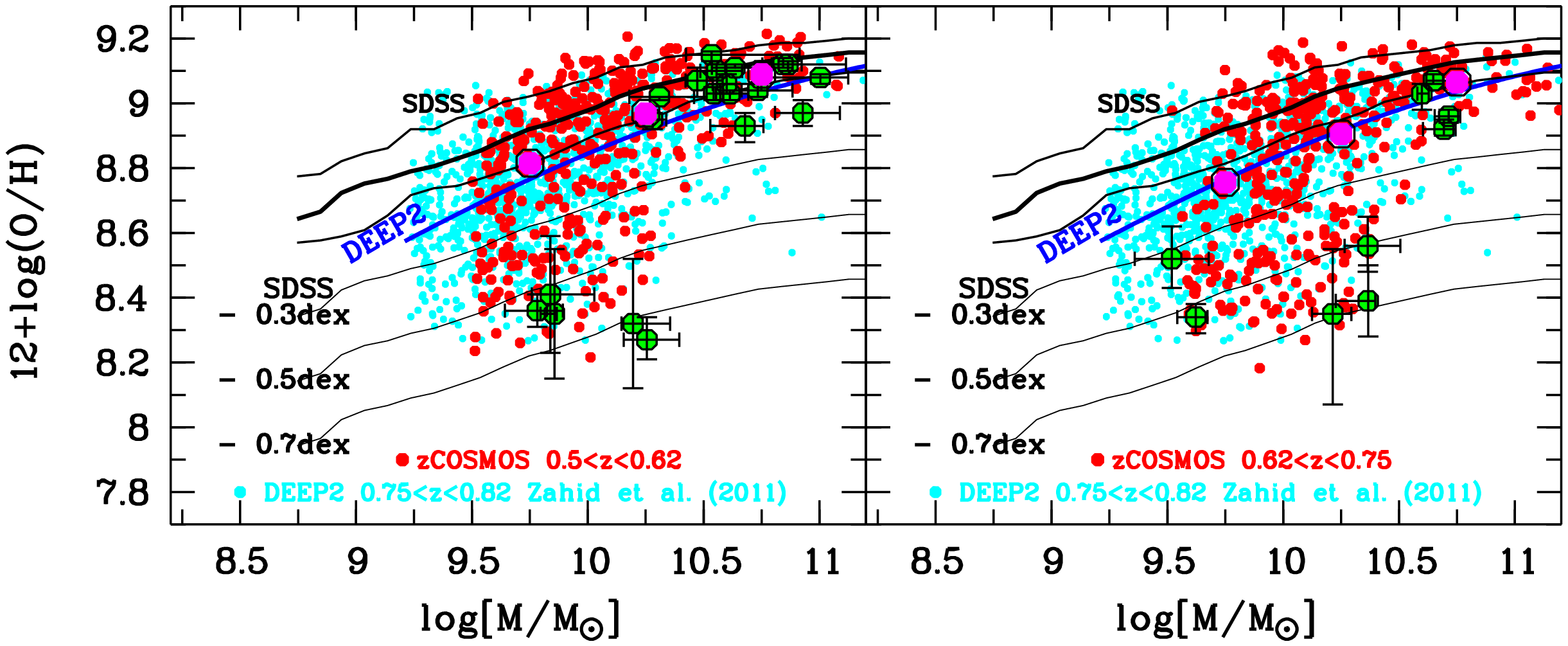}
\caption
{
\label{MZRz07} 
\footnotesize 
The MZR of zCOSMOS ELOX $0.5<z<0.75$ galaxies (red circles) and galaxies with O/H measurements based on additional ISAAC spectroscopy (green filled circles),
compared to  the local MZR of \citet{trem04} and the MZR at  $0.75<z<0.82$ found by \citet{zahid11} using the DEEP2 sample.
The individual DEEP2 data are shown as cyan symbols, while the mean DEEP2 MZR from Eq. 8 in \citet{zahid11}, converted to the KD02 calibration,
is shown as a blue line.
The mean O/H values of zCOSMOS galaxies in three mass bins are shown as big magenta filled circles.
While both zCOSMOS and DEEP2 galaxies at $z \sim 0.5-0.8$ have similar (slightly lower) metallicities at the higher masses  compared to $z \sim 0$, a population of galaxies with metallicities lower by a factor 2-3 compared to SDSS exists at the lowest masses probed by zCOSMOS and DEEP2. 
}
\end{figure*}

Fig.\,\ref{MZRz07} shows the MZR of  $0.5<z<0.75$ zCOSMOS galaxies, compared to SDSS.
For the local MZR, 
we converted the oxygen abundances of \citet{trem04} to the KD02 calibration using the \citet{keweli08} conversion, and also converted their stellar masses to a \citet{salp55} IMF.
16th and 84th percentiles, and the medians (50th percentiles) of the distribution of O/H in the respective mass bin are shown as black thick lines for SDSS, and the median SDSS MZR, 
shifted downward by 0.3, 0.5 and 0.7 dex, respectively, is shown by the three thinner black lines.
 Our ISAAC based O/H measurements for the 29 galaxies at $0.5<z<0.75$ with reliable metallicity measurements are shown as larger green filled circles, while the zCOSMOS ELOX measurements are shown as small filled red circles.

The O/H measurements from the work of \citet{zahid11} using more than 1000 galaxies from the  Deep Extragalactic Evolutionary Probe 2 (DEEP2) sample at slightly higher redshift, $0.75<z<0.82$, are shown in Fig.\,\ref{MZRz07} as cyan points. The blue line shows the mean DEEP2 MZR from Eq. 8 in \citet{zahid11}, converted to the KD02 calibration by applying the conversion given in Table\,3 of  \citet{keweli08}.
The error bars for the stellar masses of the zCOSMOS galaxies observed with ISAAC are reported in Table\,\ref{ISAACfluxes}. Stellar masses for the zCOSMOS ELOX sample (red points) and for the  DEEP2 sample (cyan points) were shown  by \citet{bolzo10} and \citet{zahid11} to be reliable within $0.2-0.3$\,dex, i.e., within a factor $1.5-2$.

About the consistency between the measurements of masses and metallicities between the SDSS and zCOSMOS sample we would like to note the following. 
Concerning the metallicities, because of the \citet{horne} method we are using, the line ratios measured for zCOSMOS galaxies are dominated by the inner regions of galaxies (with the largest surface brightness). This is roughly consistent with the measurements from SDSS spectra which are taken by placing fibers on the centers of galaxies.
Regarding the masses, Fig.\,18 of  \citet{moust13}  shows the comparison between stellar masses from SDSS  and SED fitting indicating that the two methods of estimating stellar masses are consistent within the uncertainties.

\citet{zahid11} found that high stellar mass galaxies with log(M/\msun )$ \sim 10.8$ (and higher masses) have metallicities similar to the SDSS sample, only 0.05\,dex lower, which is within the errors of the metallicity calibration conversions. For lower masses they found that the metallicity difference increases, up to $\sim 0.15$\,dex for galaxies with log(M/\msun )$\sim9.4$, indicating the low mass MZR slope is getting steeper at higher redshift compared to local galaxies.
The mean O/H values, in three mass bins, of zCOSMOS galaxies at $0.62<z<0.75$, shown as big magenta filled circles in Fig.\,\ref{MZRz07}, are in agreement with the \citet{zahid11} DEEP2 mean relation at similar (slightly higher) redshifts $0.75<z<0.82$.
This indicates that the MZR evolves to lower metallicities with increasing redshift and the low mass MZR slope is getting steeper at $z \approx 0.7$; however, the selection biases towards higher SSFRs at lower masses (especially in the higher redshift bin $0.62<z<0.75$), as shown in the panels c and d of Fig.\,\ref{SSFRMassz07}, may play a role by suggesting a larger evolution of the MZR compared to the intrinsic evolution.

While both zCOSMOS and DEEP2 galaxies at $z \sim 0.5-0.8$ have similar (slightly lower) metallicity at the higher masses  compared to $z \sim 0$, a population of galaxies with metallicities lower by a factor of $2-3$ compared to SDSS exists at the lowest masses probed by zCOSMOS and DEEP2. 
Given the metallicity uncertainties, this offset towards lower metallicities  at a given mass at $z \sim 0.5-0.8$ is in agreement, at least for the highest masses, with the 0.12\,dex value found by \citet{perez13} for the zCOSMOS sample at $0.6<z<0.8$ when restricting their sample to a $S/N>5$ selection in \Hb, but with the assumption of upper branch of \R23\, for all objects \citep[Table 3 in][]{perez13}.
On the other hand, at the lower masses we find more lower metallicity galaxies compared to the \citet{perez13} study, because we break the \R23\, degeneracy for the ELOX sample as described in Sect.\,\ref{R23degen}.
%


\begin{figure*}
\centering
\includegraphics[width=12cm,angle=270,clip=true]{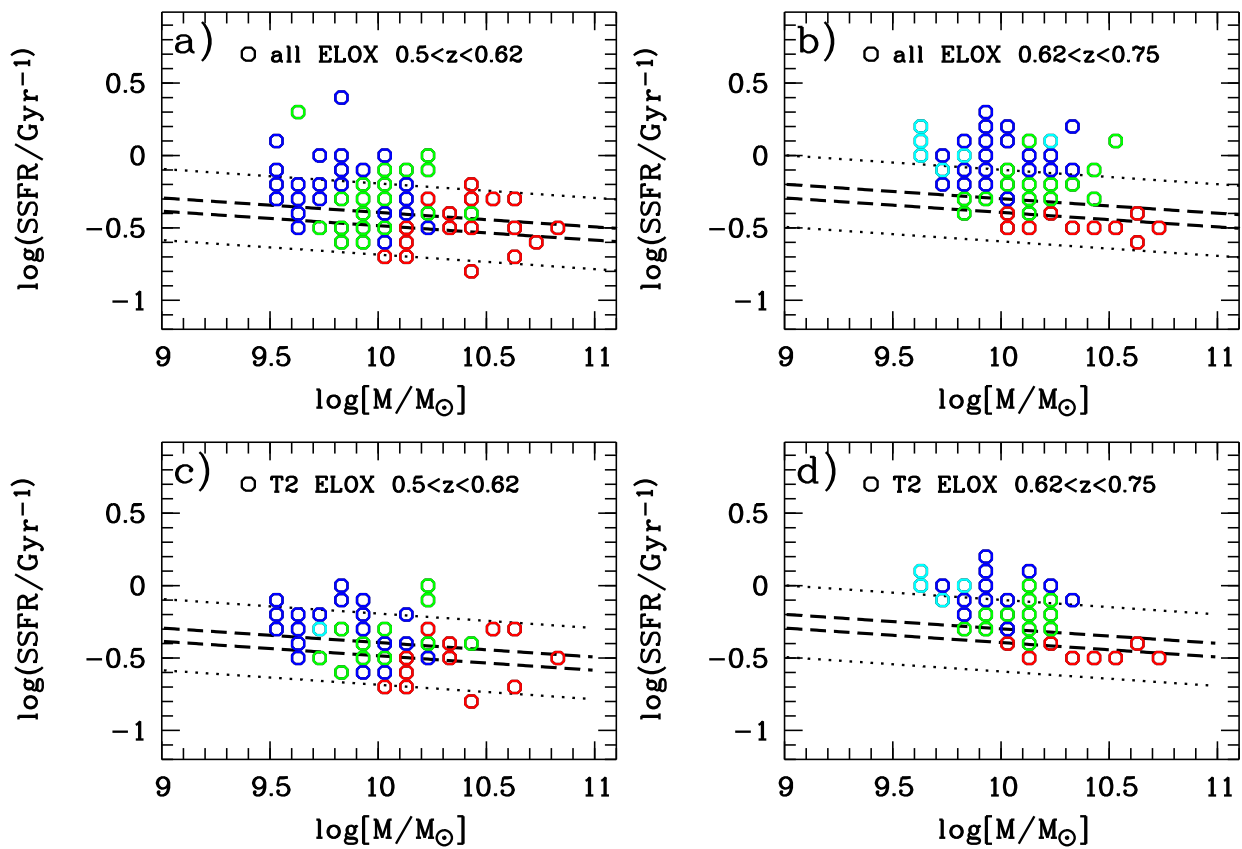}
\caption
{
\label{ZMSFR} 
\footnotesize 
The SSFR-mass relation for the ELOX galaxies at $0.5<z<0.62$ (panels a and c) and at $0.62<z<0.75$ (panels b and d), with the lower panels showing only galaxies with T2 (disk) morphologies.
Open circles depict the mean (color-coded like in Fig.\,\ref{SSFRMassz07}) metallicity data of ELOX galaxies at the position of grid points with at least three ELOX galaxies within $\pm 0.05$dex in logSFR and logM from the SSFR and mass values of the respective grid point.
It seems that the SFR is a second parameter in the MZR of the zCOSMOS sample at $0.5<z<0.75$, because galaxies with higher (S)SFRs have generally also lower metallicities.
}
\end{figure*}


\begin{figure*}
\centering
\includegraphics[width=12cm,angle=270,clip=true]{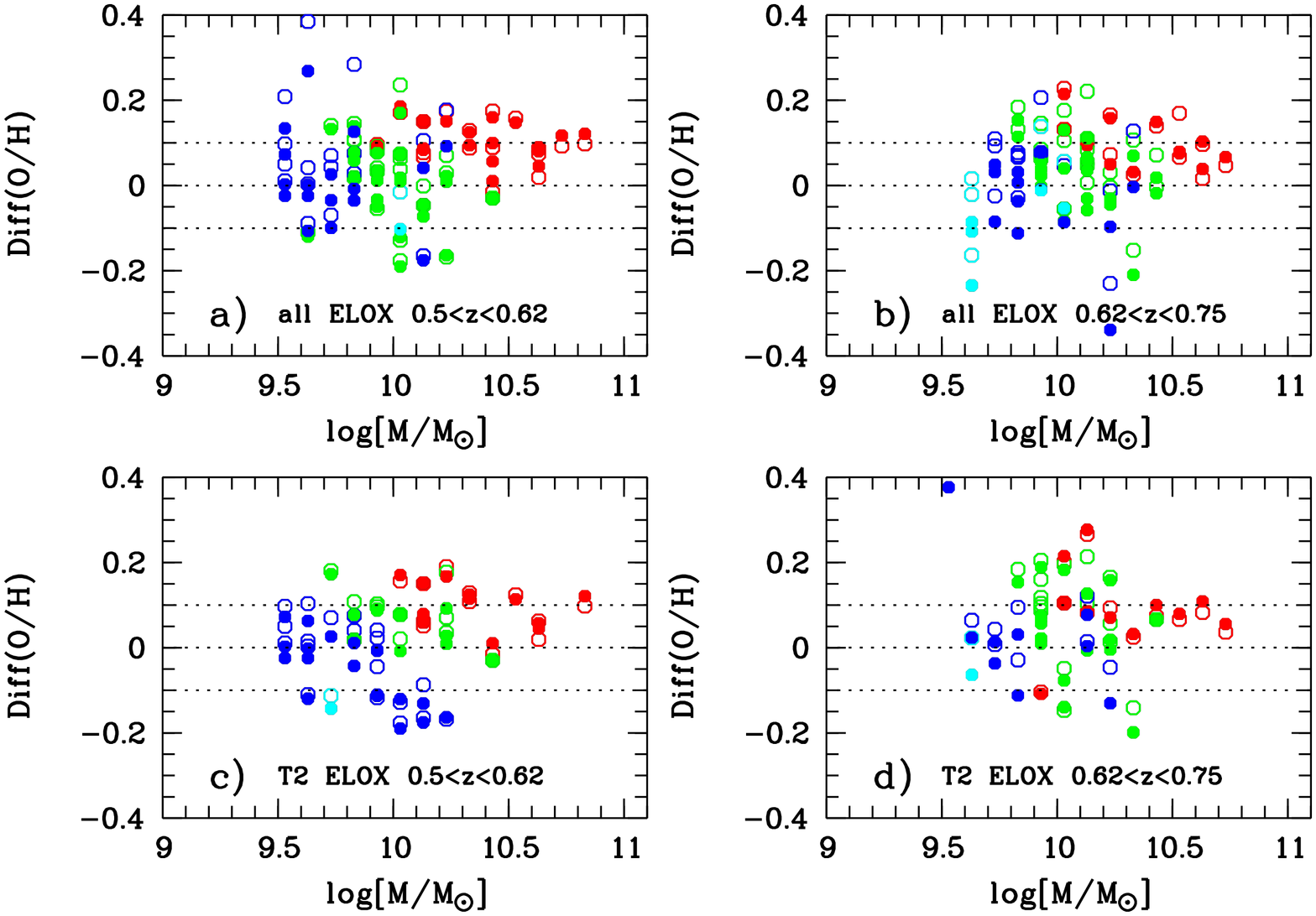}
\caption
{
\label{DiffZMSFR} 
\footnotesize 
The difference between the mean O/H at the position of grid points (color-coded like in Fig.\,\ref{SSFRMassz07}) and the expected O/H from the formulations of Li13 (open circles) and M10 (filled circles). ELOX galaxies at $0.5<z<0.62$ are shown in panels a and c, and at $0.62<z<0.75$ in panels b and d, with the lower panels showing only galaxies with T2 (disk) morphologies.
Given the fact that typical O/H uncertainties are $0.1-0.2$\,dex, the agreement between the measured and expected O/Hs is quite good, especially in panel c.
}
\end{figure*}


It should be noted that the lower number of low mass ($9.5<log(\rm{M}/\rm{M}_{\sun})<10$)  galaxies with high metallicities  at $0.62<z<0.75$ compared to $0.5<z<0.62$ is very likely due to the selection effects shown in Fig.\,\ref{SSFRMassz07} and discussed in Sect.\,\ref{SSFRMsel}. More MS galaxies at $z>0.62$ at the lower masses and lower SSFR are missed from the sample compared to $z<0.62$: the magenta symbol showing the  mean SSFR at $9.5<log(\rm{M}/\rm{M}_{\sun} \le 10$ at $z>0.62$ (panel d of Fig.\,\ref{SSFRMassz07}) lies above the MS.
This shows that at higher redshifts selection effects can produce a much steeper MZR (because of the missing low mass, low SFR, high metallicities galaxies) than an unbiased sample. Any conclusion on the evolution of the slope of the MZR with redshift must be therefore treated with caution; therefore we do not plot in Fig.\,\ref{MZRz07} the mean MZR relation of the ELOX sample, and do not discuss \emph{the exact value of the low-mass slope of the MZR} of the ELOX sample. Nevertheless, we find indication that the low-mass MZR slope is getting steeper at $z \approx 0.7$ compared to local galaxies.

The existence of $\sim 10^{10}$\msun\, $z \approx 0.7$ galaxies with metallicities lower by a factor of $2-3$ than local similar mass galaxies can be interpreted as the chemical version of galaxy downsizing \citep[cf.][]{maier06}.
''Down-sizing'' was originally introduced by \citet{cowie96} to describe the observational fact that a mass threshold decreases (in mass) with cosmic time. \citet{cowie96} showed that a threshold in mass exists below which ``forming
galaxies'' (with their timescale of formation M/SFR less than the Hubble time at that redshift)  were found and above which they were not found.
This concept has been 
generalised and used to any situation in which 
signatures of galaxy "youthfulness"  exhibit a similar mass threshold which decreases with epoch. Irregular morphologies, blue colours, high equivalent width ELs, and, as  earlier argued in \citet{maier06}, low O/Hs, have been used as  signatures of galaxy "youthfulness".
As shown in Fig.\,\ref{MZRz07} for the case of ``chemical downsizing'', the mass threshold for finding low metallicities decreases from $\sim 2\times 10^{10}$\msun\, at $z \approx 0.7$
to below $\sim 3\times 10^{9}$\msun\,
for local SDSS galaxies, which, at $\sim 3\times 10^{9}$\msun, still exhibit quite high metallicities (O/Hs larger than $\sim 8.6$).


\subsection{SFR as a second parameter in MZR, and the FMR}
\label{SFRsec}

Fig.\,\ref{SSFRMassz07} already indicates that SFR is a second parameter in the MZR of zCOSMOS galaxies at $0.5<z<0.75$, because, at a given mass, galaxies with higher (S)SFRs have generally also lower metallicities (color of the data points).
 To explore this issue further, we use Fig.\,\ref{ZMSFR}, a different version of the SSFR-mass relation for the zCOSMOS ELOX galaxies at $0.5<z<0.62$ (panels a and c) and at $0.62<z<0.75$ (panels b and d), with the lower panels showing only galaxies with T2 (disk) morphologies.
We calculated, for each point in the SSFR - mass grid, the mean values of the O/H for galaxies in the ELOX sample within $\pm 0.05$dex in log SFR and logM around the (SFR, mass) values at the respective grid point. 
We show as open circles in Fig.\ref{ZMSFR} the mean  metallicity data (color-coded like in the legend of Fig.\,\ref{SSFRMassz07}) of ELOX galaxies at the position of grid points with at least three ELOX galaxies within $\pm 0.05$dex around the (SFR, mass) values at the respective grid point.
Fig.\ref{ZMSFR} shows that, in general, at a given stellar mass, galaxies with  lower  SSFRs have higher O/Hs (red circles, O/H$>8.95$) than higher SSFR objects, which have lower O/Hs ($8.8<O/H<8.95$, green circles) and even lower $8.6<O/H<8.8$ (blue circles) for higher SSFRs. Thus, a dependency of the MZR of  $0.5<z<0.75$ zCOSMOS galaxies on SFR exists.

Fig.\,\ref{DiffZMSFR} shows the difference between the mean O/H at the position of grid points (color-coded like in Fig.\,\ref{SSFRMassz07}) and the expected O/H from the formulations of Li13 (open circles) and M10 (filled circles). 
 Most disk (T2) galaxies are  in agreement with both M10 (their Eq.2) and Li13 FMR predictions, particularly at $0.5<z<0.62$, where these disk galaxies lie mainly on the main sequence (panels c of Figs.\,\ref{ZMSFR} and \ref{DiffZMSFR}). 
Given the uncertainties in measuring metallicities of typically  $0.1-0.2$\,dex, the agreement between the measured and expected O/Hs from both Li13 and Ma10 formulations of the Z(M,SFR) is quite good.

\subsection{Comparison with other zCOSMOS work on Z(M,SFR)}

Because the absolute O/H values of the different works using different metallicity calibrations are not easily comparable, it is difficult to asses if low metallicity $logM/M_{\sun} \sim 10$ galaxies like the ones reported here were already presented in other zCOSMOS work.  As mentioned in  Sect.\,\ref{SSFRMsel},
the upper R23 branch of the O/H KD02 calibration we are using roughly corresponds to $O/H>8.6$.
\citet{cresci12} used the calibration of \citet{maiol08} with the upper R23 branch extending down to $O/H \sim 8.4$ \citep[see Fig.5 in][]{maiol08}.  
\citet{perez13}  used a different O/H calibration based or converted to the calibration of the N2$=log\NII/\Ha$\, parameter proposed by \citet{perez09}, yielding lower oxygen abundances by a factor of $2-3$ than the KD02 calibration.

The criterion used by \citet{cresci12} to select low \R23\, branch objects was $log[OIII]/[OII]>0.45$.
There are only two objects with $log[OIII]/[OII]>0.45$ in our zCOSMOS sample with ISAAC spectroscopy, and four galaxies in the zCOSMOS ELOX sample at $0.5<z<0.75$; i.e., according to the \citet{cresci12} criterion, virtually all zCOSMOS galaxies would lie on the upper \R23\, branch of their metallicity calibration.
\citet{cresci12} used a quite small subsample of zCOSMOS $z \sim 0.7 $ galaxies  with a strong S/N selection: $S/N>20$ and $S/N>8$ in \Hb\, flux for log(M/\msun )$>10$ and log(M/\msun )$<10$, respectively. This selection biases the sample towards high SFRs, and \citet{cresci12} concluded that the Z(M,SFR) is holding up to $z \sim 1$.
On the other hand, \citet{perez13} used a $S/N>2$ in the EL fluxes, i.e., including also galaxies with lower SFRs. They assumed that all galaxies are on the upper \R23\, branch, and found redshift evolution in the Z(M,SFR) relation. 

While \citet{cresci12} finds that ``lack of evolution of the FMR at $0.2 < z < 0.8$ is fully confirmed for the zCOSMOS sample'' using a sample with quite high SFRs,
it seems that a  $S/N>2$ EL flux selection like the one used by \citet{perez13} to study the Z(M,SFR) contains some galaxies
with rather low SFRs which do not follow the Z(M,SFR) relation.
The ELOX sample in this work has a $S/N>5$ selection in \Hb\, flux for the reasons given in Sect.\,\ref{ELs}, i.e., an \emph{intermediate} $S/N$ limit compared to the \citet{cresci12} and \citet{perez13} samples.
We find that most (but not all) galaxies in the zCOSMOS sample are consistent with no evolution in the Z(M,SFR), as shown in Figs.\,\ref{SSFRMassz07}, \ref{ZMSFR} and \ref{DiffZMSFR}, obtaining an \emph{intermediate} result compared to the \citet{cresci12} and \citet{perez13} studies of the Z(M,SFR) evolution. This indicates that the selection of the sample affects the dependence of the MZR on SFR; the effects of the selection of the sample should be explored in future studies by using the SSFR-mass diagram, as we did in Figs.\,\ref{SSFRMassz07}, \ref{ZMSFR} and \ref{DiffZMSFR}.


\section{Conclusions}

This study of the MZR and its dependence on SFR at intermediate redshifts is based on ISAAC NIR spectroscopy of 39 zCOSMOS galaxies. The results of the NIR spectroscopy have been used as a benchmark to calibrate the metallicities and SFRs of a much larger zCOSMOS sample of  900 galaxies, to study the Z(M,SFR) relation at $z \approx 0.7$.

The main results can be summarized as follows:

1. The zCOSMOS galaxies at $0.5<z<0.9$ with NIR spectroscopy are not dominated by AGNs.

2. We empirically found a criterion to distinguish the lower/upper branch \R23, using the position of galaxies with reliable O/Hs in the $D_{4000}$ vs. \OIIIa/\Hb\, diagram.
Our findings indicate that the low metallicity \R23\, solution should be chosen for galaxies with  $D_{4000}<1.2$ and [OIII]/H$\beta >1.8$.

3. The  MZR relation at $z \approx 0.7$ for \emph{massive}  zCOSMOS galaxies is only slightly shifted to lower metallicities compared to the SDSS sample, consistent with the findings of \citet{zahid11} at slightly higher redshift $z \sim 0.8$. 
There exists a fraction of 19\% of lower mass $9.5<log(\rm{M}/\rm{M}_{\sun})<10.3$ zCOSMOS galaxies 
which shows a larger evolution of the MZR relation, being more metal poor at a given mass by a factor of $2-3$ compared to SDSS (Fig.\,\ref{MZRz07}).
This is indication that the low-mass MZR slope is getting steeper at $z \approx 0.7$ compared to local galaxies.  

4. The appearance of $z \approx 0.7$ intermediate mass ($\sim 10^{10}$\msun) galaxies with metallicities lower by a factor of $2-3$ than local galaxies of the same mass can be interpreted as the chemical version of galaxy downsizing \citep[cf.][]{maier06}.
The mass threshold for finding low metallicities decreases from $\sim 2\times 10^{10}$\msun\, at $z \sim 0.7$
to below $\sim 3\times 10^{9}$\msun\,
for local SDSS galaxies, which, at $\sim 3\times 10^{9}$\msun, still exhibit quite high metallicities (see Fig.\,\ref{MZRz07}).

5. We find direct indications that SFR is still a second parameter in the MZR at $0.5<z<0.75$, in the sense that $0.5<z<0.75$ galaxies with higher metallicities have lower (S)SFRs (see Figs.\,\ref{SSFRMassz07} and \ref{ZMSFR}).

6. A comparison with the metallicities expected for a non-evolving FMR at $z \approx 0.7$ shows that low metallicity, $9.5<log(\rm{M}/\rm{M}_{\sun})<10.3$ galaxies as found in the zCOSMOS sample, are in agreement with the predictions of a non-evolving Z(M,SFR) (see Fig.\,\ref{SSFRMassz07}). The agreement between the measured metallicities and the predicted metallicities for a non-evolving FMR is good for the entire zCOSMOS ELOX sample (Fig.\,\ref{DiffZMSFR}). 

7. This study emphasizes that assuming the upper-branch \R23\, metallicity calibration for galaxies at intermediate redshifts would missclassify as metal-rich a substantial fraction (up to about 20\% in a given mass range) of truely low metallicity galaxies.
%


\begin{appendix}

\section{Further exploring why breaking the \R23\, degeneracy using \OIIIa/\Hb\, works}

\begin{figure}
\includegraphics[width=8cm,angle=270,clip=true]{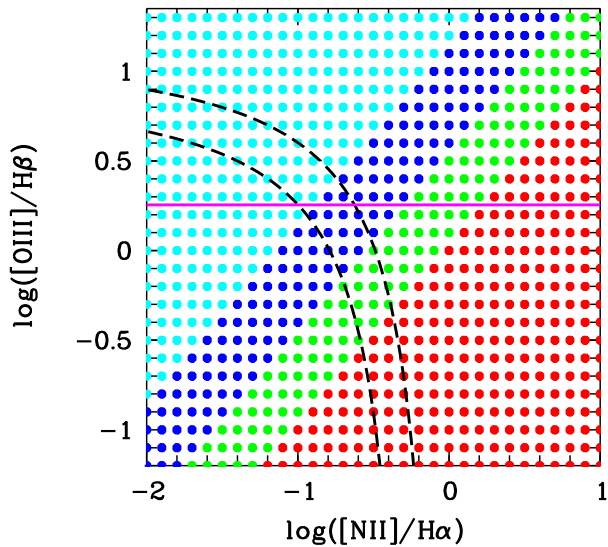}
\caption
{
\label{BPT_OIIIHb} 
\footnotesize 
\OIIIa/\Hb\, vs. \NII/\Ha\, diagram with a grid color-coded like in the legend of Fig.\,\ref{SSFRMassz07} according to metallicity.
Virtually all galaxies on the star-forming sequence (between the black dashed lines) and above the magenta line (i.e., with \OIIIa/\Hb$>1.8$) have $O/H<8.6$ on the KD02 metallicity scale (cyan symbols).}
\end{figure}

The finding in Sect.\,\ref{R23degen} regarding breaking the \R23\, degeneracy using \OIIIa/\Hb\, may appear surprising, because \OIIIa/\Hb\, has a similar behaviour as a function of metallicity like the double-valued \R23. However, some studies in the literature using the BPT \citep{bald81} diagram, e.g., Fig.\,2 in \citet{asari07} and Fig.\,1 in \citet{kewley13a}, have showed that high  \OIIIa/\Hb\, correspond to low metallicities because of the curved shape of the star-forming sequence  in the BPT diagram. 
We produced a BPT-like diagram, Fig.\,\ref{BPT_OIIIHb}, showing as a grid, color-coded like in the legend of Fig.\,\ref{SSFRMassz07}, the expected O/H values using Eq.\,4 in \citet{kewley13a},  which corresponds to the \citet{petpag04} O3N2 calibration for the KD02 metallicity scale. Because emission-line galaxies lie along the star-forming sequence (with rough boundaries indicated by the dashed black lines), virtually all galaxies  with \OIIIa/\Hb$>1.8$ values (above the magenta line), have $O/H<8.6$ on the KD02 metallicity scale (cyan colors). This is in agreement with the empirically finding of Fig.\,\ref{CaliOH} for zCOSMOS galaxies. We explore this issue further in the Clash paper mentioned in  Sect.\,\ref{R23degen} (Maier et al. 2015, im preparation), where we also show that Clash emission line galaxies at intermediate redshifts lie  on the star-forming sequence between the dashed black lines shown in Fig.\,\ref{BPT_OIIIHb}. It should be also noted that Fig.\,\ref{BPT_OIIIHb} also gives a hint why the O3N2 calibration works better at higher redshifts than the N2 calibration of \citet{petpag04}, as claimed, e.g., by \citet{liu08} and \citet{steidel14}: if the star-forming sequence moves at higher redshift diagonally to higher \OIIIa/\Hb\, and \NII/\Ha\, values \citep[as shown, e.g., in][]{kewley13b}, the O3N2 calibration still works (color of the the grid points stays the same), while the N2 calibration gives too high values for the metallicity compared to local samples.  We will discuss these findings further in our paper mentioned above \citep[Maier et al. 2015, in preparation, first results shown in][]{kuchn14}.

\end{appendix}


\begin{acknowledgements}
We are grateful the anonymous referee for his/her suggestions which have improved the clarity of the paper. We greatly acknowledge the contributions of the entire COSMOS collaboration, consisting of more than 70 scientists.
\end{acknowledgements}


\end{document}